\documentclass[12pt,preprint,psfig]{aastex}
\shorttitle{IRAC Observations of CO $J=4\rightarrow3$ Cloud in the
30 Doradus} \shortauthors{Kim et al.}

\slugcomment{Submitted to ApJ}

\begin{document}

\title{IRAC Observations of CO $J=4\rightarrow3$ High-Velocity Cloud in the 30 Doradus Complex in the Large Magellanic Cloud}
\author{Hak-Sub Kim\altaffilmark{1,5}, Sungeun Kim\altaffilmark{1*,2}, Jih-Yong Bak\altaffilmark{1}, Mario Garcia\altaffilmark{1},
Bernard Brandl\altaffilmark{3}, Kecheng Xiao\altaffilmark{2},
Wilfred Walsh\altaffilmark{2}, R. Chris Smith\altaffilmark{4}, \&
Soyoung Youn\altaffilmark{1}} \altaffiltext{1}{ARCSEC, Department
of Astronomy \& Space Science, Sejong University, KwangJin-gu,
KunJa-dong 98, Seoul, 143-747, Korea; e-mail: sek@sejong.ac.kr}
\altaffiltext{3}{Harvard-Smithsonian Center for Astrophysics, 60
Garden St., MS-12, Cambridge, MA 02138, USA; e-mail:
skim@cfa.harvard.edu} \altaffiltext{4}{Leiden Observatory,
Sterrewacht Leiden, P.O. Box 9513, Neils Bohrweg2, RA Leiden,
2300, Netherlands} \altaffiltext{5}{NOAO, 950 N. Cherry Ave.,
Tucson, AZ85719, USA}
\altaffiltext{5}{Yonsei University, Astronomy Department,
Shinchon, Seodaemungu, Seoul, Republic of Korea}
\altaffiltext{*}{Corresponding Author}

\begin{abstract}

We present the results of $^{12}$CO $J=2\rightarrow1$ observations
of the X-ray bright giant shell complex 30 Doradus in the Large
Magellanic Cloud (LMC) using the Antarctic Submillimeter Telescope
and Remote Observatory (AST/RO). This is the one of the largest
H\,II complexes in the Local Group.
We compare the $^{12}$CO $J=2\rightarrow1$ observations against
previously taken $^{12}$CO $J=4\rightarrow3$ observations and analyze
the spatial distribution of young stellar objects (YSO's) within the
cloud using the Spitzer IRAC observations of the 30 Doradus complex.
Both peaks of $^{12}$CO $J=2\rightarrow1$ and $J=4\rightarrow3$ emitting
clouds coincide with the densest region of the filaments where multiple
shells are colliding.
We find that the YSO's are clustered in the southern ridge of the warm
and dense molecular gas clouds traced by $^{12}$CO $J=4\rightarrow3$,
indicating a filamentary structure of star formation throughout the 30
Doradus.
We also find that some of Class I YSO's candidates which are likely to be
Associated with a high-velocity component of $^{12}$CO $J=4\rightarrow3$
emitting clouds are present.
This is a bona fide place where the triggered star formation had happened
and newly formed stars may have produced such a high-velocity outflow
interacting with the surrounding molecular cloud material.
\end{abstract}

\section{Introduction}
\label{s:intro}

The Large Magellanic Cloud (LMC) provides an excellent opportunity
to study the effects of radiation from young star objects (YSO's)
on their multi-phase interstellar medium (ISM). The LMC has a
27-degree incline to the line-of-sight and has small foreground
and internal extinctions, making it possible to map the gas and
dust of the ISM in the LMC without confusion with extraneous
matter along the line-of-sight. The LMC stars are at a common
distance from the Earth, unlike stars inside of the Milky Way's
galactic plane, but are still close enough that individual stars
and their stellar ejecta can be studied in great detail. We have
chosen the 30 Doradus nebula inside of the LMC in particular given
that Kim et al. (2005) detected a high-velocity molecular emission
associated with the H\,II and H\,I shells, indicating that this is a
very active region of star formation. Additionally, the LMC provides
an excellent opportunity to study the effects of different UV radiation
from stars on their environments in the multi-phase ISM, as well as to
apply their observed mechanisms to studies of the early evolution of
high-redshift, metal-poor galaxies. CO is the chosen species to trace the
general distribution of molecular gas because it is the most abundant
observable tracer of H$_2$ and is sensitive to a density regime prevalent
in the diffuse molecular gas surrounding dark clouds, molecular cores, and
star-forming regions. For Galactic work, the $^{12}$CO line is usually
saturated and an isotope replacement molecule such as $^{13}$CO is needed
to obtain better information on column densities and cloud sizes. In the LMC,
and therefore in 30 Doradus as well, the diffuse CO component is weaker due
to lower abundance and higher photodissociation rates, so $^{12}$CO is a
less-biased molecular tracer which leads an assumption that the cloud is
virialized and a reliable estimate of X conversion factor (Dickman 1978).

The present study aimed to characterize the relationship between the warm
and dense core of the molecular clouds traced with $^{12}$CO $J=4\rightarrow3$
and the distribution of young stars across the clouds in the 30 Doradus nebula.
The mid-$J$ CO emissions in the 30 Doradus region have recently been observed
with the Antarctic Submillimeter Telescope and Remote Observatory (AST/RO) and
$^{12}$CO $J=2\rightarrow1$ emission from the 30 Doradus reveal the extended
structure beyond the $^{12}$CO $J=4\rightarrow3$ emitting region. With the
advent of the Spitzer Space Telescope, the distribution of young stellar objects
across the region can be mapped with the Spitzer/IRAC photometry (Meixner et al.
2006). Jones et al. (2005) provide comprehensive study of protostars in the giant
H\,II region complex N159 in the LMC with the Spitzer IRAC observations. In this
paper, color-color and color-magnitude diagram for Spitzer/IRAC sources have been
used to study the nature of distributed populations and their locations in the
molecular clouds in the 30 Doradus in conjunction with the mid-$J$ CO emissions
observed with the AST/RO.

\section{Observations}

The observations of $^{12}$CO $J=2\rightarrow1$ emission lines were performed
during the austral winter season of 2002 and 2003 at the Antarctic Submillimeter
Telescope and Remote Observatory (AST/RO), a 1.7-m diameter, offset Gregorian
telescope. AST/RO is located at 2847 m altitude at the Amundsen-Scott South
Pole Station and capable of observing at wavelengths between 200 $\mu$m and 1.3
mm (Stark et al. 2001). This site is very dry, so it is therefore a very good
position from which to make submillimeter observations (Chamberlin, Lane, \&
Stark 1997). Since the observations of $^{12}$CO $J=4\rightarrow3$ emission
line are described in Kim et al. (2005), we mention only the observations of
$^{12}$CO $J=2\rightarrow1$ emission taken with the AST/RO in this paper.

Emission from the $^{12}$CO $J=2\rightarrow1$ line was mapped over a 30$'$
$\times$ 30$'$ region centered on R.A.=$5^{\rm h}38^{\rm m}40^{\rm s}$,
Dec.=-69$^{\circ}06'28''$ (J2000) with 90$''$ spacing and a beam size of about
180$''$. The position switching mode was used and observation time at each
position was typically 7 minutes and 20 seconds. The receiver used was a 230
GHz superconductor-insulator-superconductor (SIS) waveguide receiver with
79$-$90 K double-sideband (DSB) noise temperature. Two acousto-optical
spectrometers (Schieder, Tolls, and Winnewisser 1989) were used as backends.
AST/RO had four available acousto-optical spectrometers (AOSs). An array AOS
having four low-resolution spectrometer (LRS) channels with a bandwidth of
1 GHz (bandpass 1.6$-$2.6 GHz) and a resolution of ~1 MHz corresponding to a
velocity resolution of ~1.3 km/s has been used for 230 GHz observation. LRS
has 2048 channels. AST/RO suffers pointing errors of the order of $1\arcmin$,
and the beam sizes (FWHM) were $170$--$190\arcsec$ at 230 GHz.
Atmosphere-corrected system temperatures ranged from 309 to 364 K at 230 GHz.

The standard chopper wheel calibration technique was employed, implemented at
AST/RO by way of regular (every few minutes) observations of the sky and two
blackbody loads of known temperature (Stark et al. 2001). Atmospheric
transmission was monitored by regular skydips, and known, bright sources
were observed every few hours to further check calibration and pointing. At
periodic intervals and after tuning, the receivers were manually calibrated
against a liquid-nitrogen-temperature load and the two blackbody loads at
ambient temperature and about 100 K. The latter process also corrects for the
dark current of the AOS optical CCDs. The intensity calibration errors became
as large as $\pm15$\% during poor weather periods.

Once taken, the data in this survey were reduced using the COMB data reduction
package. After elimination of scans deemed faulty for various instrumental or
weather-related reasons ($\sim 10\%$ of the total dataset), the data were
Fourier transformed and the second highest amplitude channel was zeroed out,
so well-defined baseline component was removed. Then, linear
baselines were removed from the spectra. This is shown in Figure 1. The Fourier
transforms are done using fast Fourier transform techniques and the spectrum
is padded with zeros on the right to the next biggest power of 2 (e.g. 2048
channels). After a Fourier transform of the plotted channels of the original
full spectrum were done and then selected high amplitude channel was removed,
the data were re-transformed and inserted back into the original spectrum. Since
the operations are all done in double precision, there is no loss of accuracy.

%FIGURE 1: SPECTRA
\dataset[ADS/Sa.Spitzer#0004379904]{In order to detect young
stellar objects (YSOs), we used the Spitzer archival data of 30 Doradus nebula
(AORKEY 4379904) observed with the Infrared Array Camera (IRAC). IRAC data
consist of four broadband images, each of which has a central wavelength of
3.6, 4.5, 5.8, and 8.0$\mu$m (Fazio et al. 2004).} First, we performed an
array-location-dependent photometric correction on the Basic Calibrated Data
(BCD) processed by the IRAC pipeline (version S13.2.0). This correction is
required because the pipeline flat-fielding based on the zodiacal background
is not appropriate for point source analysis. In the next step, we corrected
``muxbleed'' and ``column pulldown'' effects, which are image
artifacts that appear around bright sources, and matched
background levels of overlapping frames. Finally, mosaicked images
were made from these corrected images using MOPEX mosaicker which
is provided by the Spitzer Science Center (SSC).

Source extraction and aperture photometry were performed using the
Spitzer Astronomical Point Source Extractor(APEX) with 5 pixel
aperture radius and 5-10 pixel background annulus. Pixel phase
corrections, which correct the dependence of photometry on the
location of a source within its peak pixel, were applied on the
photometry of channel 1, and aperture corrections were applied on
all channels. Then, we converted flux densities into magnitudes
using the IRAC zero-magnitude flux densities. All the values used
for corrections and calculations were obtained from the IRAC Data
Handbook version 3.0 (http://ssc.spitzer.caltech.edu/irac/dh/).

\section{Results}
\label{s:results}

%COMPARISON OF 12CO 1-0 TO 12CO 4-3
We have observed $^{12}$CO $J=2\rightarrow1$ line emission
associated with the 30 Doradus nebula in the LMC (Figure 2).
The peak of $^{12}$CO $J=2\rightarrow1$ emission is only 40$''$ away
from the peak of $^{12}$CO $J=4\rightarrow3$ emission. Including
the pointing error and the beam size of $^{12}$CO $J=2\rightarrow1$
emission, the peak in $^{12}$CO $J=4\rightarrow3$ emission is very
close to that in $^{12}$CO $J=2\rightarrow1$ emission. Both peaks are
not associated with the R136 cluster. Instead they are surrounded
by R145 (WN6h), R139 (O6Iaf/WN)/R140 (WN6), and R144 (WN6h) multiple
system north of the R136 cluster (see Figure 2).

%FIGURE 2a: CO 2-1 Figure 2b: R136, etc. marked.
%FIGURE 3: X-ray image
The peaks of $^{12}$CO $J=4\rightarrow3$ and $^{12}$CO
$J=2\rightarrow1$ emissions are spatially distanced from the peaks
of H$\alpha$ and X-ray emissions as seen in Figure 2 and Figure 3
as it is expected. The apparent H\,II regions and the bright X-ray
counterparts in this region indicate that both the fully ionized
regions close to the massive stars, with the related ionizing
radiation escaping from the H\,II regions, heat the H molecules
and keep them dissociated. The morphology of H$\alpha$ emission is
almost identical to that of 8 $\mu$m emission (Figure 4), tracer
of Polycyclic Aromatic Hydrocarbon (PAH) emission, indicating that
the PAH is only partially destroyed in the H\,II region. This is
in contrast to the pretty good correlation shown in the SAGE study
by Meixner et al. (2006) and SINGS study by Regan et al. (2006).
%FIGURE 4: 8um+STELLAR CANDIDATES
High-velocity $^{12}$CO $J=4\rightarrow3$ emitting gas detected in this
region (Kim et al. 2005; Kim 2006) and the morphology of $^{12}$CO
$J=4\rightarrow3$ emission indicate that the bulk of the CO emission
must come from the entrainment of ambient atomic and molecular cloud
gas by the stellar winds colliding with one another seen in the H$\alpha$
image (Figure 2). Using the best estimate of the luminosity mass for
$^{12}$CO $J=4\rightarrow3$ outflow (Kim et al. 2005), 2.6$\pm$1.6$\times$
10$^4$ $M_\odot$, and the dynamical age of the shells, 10 Myr, we
estimate the rate of ambient gas entrainment to be $M_{flow}$=2.6
$\times$ 10$^{-3}$ $M_\odot$ yr$^{-1}$ with the product of preshock
density and velocity, approximately $nv$ $\sim$ 18 cm$^{-3}$ km s$^{-1}$.
The momentum injection rate is $P_{flow}=1.3\times10^{-1}M_\odot$
km s$^{-1}$yr$^{-1}$ and the mechanical luminosity is to be
$E_{flow}$=5.4 $\times$ 10$^2$ $L_\odot$.

We detected the near-IR sources associated with the molecular
cloud core in the $^{12}$CO $J=2\rightarrow1$ and $^{12}$CO
$J=4\rightarrow3$ emitting clouds in the 30 Doradus nebula by
utilizing Spitzer GTO IRAC observations. We identified YSO
candidates among these sources from the color-color diagram of
3.6 $\mu$m $-$ 4.5 $\mu$m versus 5.8 $\mu$m $-$ 8.0 $\mu$m (Figure
5a), and classified them into three groups based on the criterion
suggested by Allen et al. (2004). Since the reddening vector for
30 Doradus is about 1.3 mag, it can only change the colors of
[3.6 $\mu$m]$-$[4.5 $\mu$m]=0.01 mag and [5.8 $\mu$m]$-$[8.0 $\mu$m]=0.02
mag (Rieke and Lebofsky 1985). This is too small to affect the
classification.
These classes show the evolutionary sequence from Class \textrm{I}
to Class \textrm{III}. Class \textrm{I} objects are surrounded by
infalling envelopes, Class \textrm{II} objects have accretion
disks, and Class \textrm{I}/\textrm{II} objects are in an
intermediate state between Classes \textrm{I} and
\textrm{II} (Kenyon \& Hartmann 1995, Megeath et al. 2004).
They are denoted by open circles (Class \textrm{I}), squares (Class
\textrm{II}), and diamonds (Class \textrm{I}/Class \textrm{II})
respectively in Figure 5. Class \textrm{III} objects were excluded
from the classification because they have the SEDs of stellar
photospheres and cannot be distinguished from foreground or
background stars. Since the brightness of 8 $\mu$m emission can be
affected by PAH emission arising
from the 30 Doradus nebula, we present the color-color diagram of
3.6 $\mu$m $-$ 4.5 $\mu$m versus 4.5 $\mu$m $-$ 5.8 $\mu$m in
Figure 6. It confirms that the classified objects in Figure 5a fall
in a relatively well-defined region of the color$-$color diagram in
Figure 6. The distinction is clear. The only problem is that the
30Dor-09 object identified as Class \textrm{II} in Figure 5a shows
much redder [4.5 $\mu$m]$-$[5.8 $\mu$m] color $\sim$ 1.7. This object
was identified as M red supergiant star, Dor IR 10 by McGregor \& Hyland
(1981) using the $JHK$ photometry and IR spectroscopy. They also found
11 other M red supergiants in the 30 Doradus, but we could not find any
corresponding sources associated with our classified YSO's catalog.
Previous study by Brandner et al. (2001) reports 20 Class I protostars
and Herbig Ae/Be candidates near R136 cluster in the 30 Doradus complex
using $JHK$ color-color diagram. 8 of these sources could be matched
with the YSO population traced by present IRAC data analysis.
Figure 5b shows the location of these objects on the color-magnitude
diagram of 3.6 $\mu$m $-$ 8.0 $\mu$m versus 8.0 $\mu$m using the
above symbols. Class \textrm{I} objects show strong 3.6 $\mu$m
$-$ 8.0 $\mu$m color excess with color indices of approximately 4,
while most of Class \textrm{II} objects show relatively weak 3.6
$\mu$m $-$ 8.0 $\mu$m color excess with indices from 1 to 2.

%FIGURE 5a; color-color diagram, 8um
%Figure 5b; color-magnitude diagram
%Figure 6; color-color diagram
%MENTION ON THE BACKGROUND GALAXIES

\section{Discussion}
%\subsection{YSO Candidate Detections}

An HI aperture synthesis mosaicked map of the LMC with a spatial
resolution of 50$''$, created by combining 1344 separate pointings
of the Australia Telescope Compact Array (ATCA), shows an overall
clumpy HI distribution featuring holes, shells, loops, filaments,
and bubbles (Kim et al. 2003; Kim et al. 1998; Kim et al. 1999).
The HI supergiant shells occupy a large volume of the ISM and a
large number of giant shells, contained within each supergiant,
are colliding with one another. The many shells observed often
overlap and are interacting with one another, such as in the
vicinity of the 30 Doradus complex where very active star
formation has been happening simultaneously in many different
centers.

A number of the smaller shells formed on the rims of supergiant
shells and are found in regions of very active star formation in
the LMC (Kim et al. 1999). Recent submillimeter observations of
the LMC supershells and superbubbles in the LMC using the
Antarctic Submillimeter Telescope and Remote Observatory (AST/RO)
aimed to look for the place where the triggered star formation had
happened and found the $^{12}$CO $J=4\rightarrow3$ emission toward
the rim of the expanding giant H\,II complex in the 30 Doradus
(Kim et al. 2005). They detected a possible high-velocity
molecular emission associated with the H\,II and H\,I shells in
the 30 Doradus nebula. This observational fact has been understood
as the result of self-propagating star formation, where
gravitational instabilities in the swept-up material of the
supergiant shell caused fragmentation and a new round of star
formation. Newly formed stars may have produced such a high
velocity molecular outflows interacting with the surrounding
molecular cloud material. Outflows and jets are commonly seen in
the ambient medium of young stellar objects.

It is commonly accepted that stars form in molecular clouds by the
gravitational collapse of dense gas. Both low and high mass young
stellar objects are embedded in the parent molecular and cause the
heating of the surrounding gas. Often outflows from young stellar
objects accelerate gas and are observed as molecular outflows. The
radiative heating and magnetic fields are likely to be the main
energy sources driving outflows for both low and high mass young
stellar objects. Using the recent Spitzer IRAC observations of 30
Doradus complex, we examine young stellar objects traced by IRAC data
which might be associated with the high-velocity components of $^{12}$CO
$J=4\rightarrow3$ emitting clouds.

A total of 41 YSO candidates, which fall close to the core of the
$^{12}$CO $J=2\rightarrow1$ and $^{12}$CO $J=4\rightarrow3$
emitting clouds, were detected and classified from the IRAC survey
of 30 Doradus . We present our photometry results and classes in
Table 1. Figure 4 presents the spatial distribution of detected
YSO candidates towards the relatively dense molecular cloud core found
in the $^{12}$CO $J=4\rightarrow3$ and $^{12}$CO $J=2\rightarrow1$ emitting
gas. A strong association is not found between the location of YSO
candidates and of dense molecular cloud core as seen in Figure 4.
However, Class \textrm{I} candidates which are thought to be
protostars surrounded by dusty infalling envelopes and exhibit
significant near-infrared extinction from their envelopes are very close to
the peaks of $^{12}$CO $J=4\rightarrow3$ and $^{12}$CO $J=2\rightarrow1$
emissions. Conspicuously, Class I YSO 30 Dor-30 is detected near the
high-velocity component of $^{12}$CO $J=4\rightarrow3$ emission.
13 YSO candidates are concentrated in the southern region of the
$^{12}$CO $J=2\rightarrow1$ emitting cloud, about 30 pc away from
the molecular cloud core. Association densities are likely to be
as low as 10$^{-4}$ stars pc$^{-3}$ at the distance of the LMC
(Feast 1991) and correspond to the number density of molecular
hydrogen $n$($H_2$) of 10$^4$ molecules cm$^{-3}$ (Kim 2006).
It is notable that the relationship between the number of YSO candidates
And their radial distribution from the core (Figure 4) reveals that
more recently formed stars (Class \textrm{I} objects) are distributed
more closely to the molecular cloud core.

Figure 7 presents an analysis that measures the distance of the
YSO's to the core of the CO cloud. The method performed was
similar to one present in the paper by Teixeira et al. (2006),
except that the measured variable is different. We utilized a
Monte Carlo simulation to analyze the difference between the
observed spatial distribution and a randomly generated star field.
We created 10,000 star fields with an equal number of randomly and
uniformly placed YSO's in an equal observing area. The average
distance to the cloud core of these objects was calculated and
presented in a histogram to be compared with the distances of the
observed YSO's. These comparisons show that there is a structural
arrangement of YSO's with respect to the core of the CO cloud in the
30 Doradus Nebula, likely to be a filamentary distribution. In
the total distribution of all classes, we find that there is a
probability of 0.15\% of having more than 9 objects placed between
200 and 300 arcseconds from the core. This seems to be mostly due
to the Class I YSO's given that there is a 0.02\% probability of
finding more than six Class I YSO's at that same distance. We can
see that Class I/II YSO's are absent at this distance, but using
the next bin (300-400 arcsec) we find that at its peak the Class
I/II YSO's have a probability of 3.41\% of having more than 4
objects. Class II objects also follow the general distribution of
the other two previous classes, but are too few and therefore
tests are inconclusive.

\section{Conclusion}

AST/RO observations have revealed that the peaks of $^{12}$CO $J=2\rightarrow1$
and $^{12}$CO $J=4\rightarrow3$ emission are located at the rims of colliding
H\,II shells (Chu and Kennicutt 1994) in the very active star-forming region in
the 30 Doradus complex in the LMC. Previous studies by Kim et al. (2005) show
that $^{12}$CO $J=4\rightarrow3$ emission has a high-velocity component corresponding
to the peak of the emission. This region must be the place where the triggered star
formation had happened since newly formed stars may have produced such a high-velocity
molecular outflow interacting with the surrounding molecular cloud material. Using the
recent Spitzer IRAC observations of 30 Doradus complex, we find that Class I YSO's
candidates are prone to be associated with the high-velocity component of
$^{12}$CO $J=4\rightarrow3$ emitting clouds. We also discover that the YSO's are
clustered in the southern ridge of the relatively warm and dense molecular cloud cores
traced by $^{12}$CO $J=4\rightarrow3$, indicating a filamentary structure of star
formation.

\acknowledgments \label{s:ack}
We thank A. A. Stark (AST/RO P.I.) and A.P. Lane for their support of
this project and helpful discussion; C. Walker and his SORAL receiver
group at the U. of Arizona; J. Kooi and R. Chamberlin of Caltech, G.
Wright of Antiope Associates, and K. Jacobs of U. K\"{o}ln for their
work on the instrumentation; R. Schieder, J. Stutzki, and colleagues
at U. K\"{o}ln for their AOSs. We thank Leisa Townsley for her
providing us X-ray image and Lori E. Allen for her helpful comments.
We thank anonymous referee for invaluable comments. This research was
in part supported by NSF grant number OPP-0126090. SK was supported in
part by Korea Science \& Engineering Foundation (KOSEF) under a cooperative
agreement with the Astrophysical Research Center of the Structure and
Evolution of the Cosmos (ARCSEC). This work is based in part on
observations made with the Spitzer Space Telescope, which is operated
by the Jet Propulsion Laboratory, California Institute of Technology
under a contract with NASA.

\newpage
\begin{deluxetable}{lccrrrrc}
\tabletypesize{\scriptsize}
%\rotate
\tablecolumns{8}
\tablecaption{YSO candidates near the 230 GHz $^{12}$CO $J=2\rightarrow1$ emission core in 30 Doradus complex}
\label{tbl-1}
\tablewidth{0pt}

\tablehead{\colhead{ID} &\colhead{RA (J2000)} &\colhead{DEC (J2000)} &\colhead{3.6 $\mu$m} &\colhead{4.5 $\mu$m} &\colhead{5.8 $\mu$m} &\colhead{8.0 $\mu$m} &\colhead{Class}}

\startdata

30Dor-01 & 05 37 42.70 & -69 09 43.20 & 12.38 $\pm$ 0.02 & 12.17 $\pm$ 0.03 &  9.89 $\pm$ 0.02 &  8.19 $\pm$ 0.02 & I/II \\

30Dor-02 & 05 37 49.50 & -69 09 54.20 & 11.51 $\pm$ 0.02 & 11.82 $\pm$ 0.03 & 10.46 $\pm$ 0.02 &  9.19 $\pm$ 0.02 & I/II \\

30Dor-03 & 05 37 50.50 & -69 04 01.80 &  9.25 $\pm$ 0.02 &  8.63 $\pm$ 0.03 &  8.01 $\pm$ 0.02 &  7.47 $\pm$ 0.02 &   II \\

30Dor-04 & 05 37 51.10 & -69 09 33.70 & 12.67 $\pm$ 0.02 & 12.46 $\pm$ 0.03 & 11.68 $\pm$ 0.02 &  9.73 $\pm$ 0.02 & I/II \\

30Dor-05 & 05 38 05.70 & -69 09 09.50 & 11.06 $\pm$ 0.02 & 10.35 $\pm$ 0.03 &  8.92 $\pm$ 0.02 &  7.13 $\pm$ 0.02 &    I \\

30Dor-06 & 05 38 16.70 & -69 04 13.60 &  8.35 $\pm$ 0.02 &  8.22 $\pm$ 0.03 &  7.91 $\pm$ 0.02 &  7.50 $\pm$ 0.02 &   II \\

30Dor-07 & 05 38 17.00 & -69 04 00.60 &  8.79 $\pm$ 0.02 &  8.62 $\pm$ 0.03 &  8.41 $\pm$ 0.02 &  7.77 $\pm$ 0.02 &   II \\

30Dor-08 & 05 38 20.00 & -69 06 42.40 & 12.60 $\pm$ 0.02 & 12.89 $\pm$ 0.03 &  9.59 $\pm$ 0.02 &  7.84 $\pm$ 0.02 & I/II \\

30Dor-09 & 05 38 27.40 & -69 08 09.00 & 12.99 $\pm$ 0.02 & 12.27 $\pm$ 0.03 & 10.59 $\pm$ 0.02 &  9.83 $\pm$ 0.02 &   II (?) \\

30Dor-10 & 05 38 30.00 & -69 05 37.40 & 10.91 $\pm$ 0.02 & 10.05 $\pm$ 0.03 &  9.15 $\pm$ 0.02 &  6.87 $\pm$ 0.02 &    I \\

30Dor-11 & 05 38 30.00 & -68 59 33.00 & 12.27 $\pm$ 0.02 & 12.45 $\pm$ 0.03 & 11.20 $\pm$ 0.02 &  9.82 $\pm$ 0.02 & I/II \\

30Dor-12 & 05 38 30.10 & -69 06 25.60 & 11.43 $\pm$ 0.02 & 11.18 $\pm$ 0.03 &  8.65 $\pm$ 0.02 &  6.87 $\pm$ 0.02 & I/II \\

30Dor-13 & 05 38 31.60 & -69 02 13.80 &  9.78 $\pm$ 0.02 &  9.03 $\pm$ 0.03 &  7.06 $\pm$ 0.02 &  5.23 $\pm$ 0.02 &    I \\

30Dor-14 & 05 38 36.40 & -69 06 21.00 & 10.63 $\pm$ 0.02 &  9.28 $\pm$ 0.03 &  8.53 $\pm$ 0.02 &  6.88 $\pm$ 0.02 &    I\tablenotemark{*}\\

30Dor-15 & 05 38 36.60 & -69 05 24.30 & 10.74 $\pm$ 0.02 & 10.42 $\pm$ 0.03 &  8.29 $\pm$ 0.02 &  6.98 $\pm$ 0.02 & I/II \\

30Dor-16 & 05 38 38.60 & -69 06 12.50 & 13.29 $\pm$ 0.03 & 11.08 $\pm$ 0.03 &  9.46 $\pm$ 0.02 &  7.02 $\pm$ 0.02 &    I\tablenotemark{*}\\

30Dor-17 & 05 38 38.90 & -69 06 49.30 & 11.08 $\pm$ 0.02 & 10.42 $\pm$ 0.03 &  9.64 $\pm$ 0.02 &  8.75 $\pm$ 0.02 &   II \\

30Dor-18 & 05 38 39.30 & -69 05 52.20 & 11.21 $\pm$ 0.02 & 10.28 $\pm$ 0.03 &  8.81 $\pm$ 0.02 &  6.87 $\pm$ 0.02 &    I\tablenotemark{*}\\

30Dor-19 & 05 38 39.60 & -69 09 57.20 & 11.76 $\pm$ 0.02 & 11.06 $\pm$ 0.03 &  9.40 $\pm$ 0.02 &  7.88 $\pm$ 0.02 &    I \\

30Dor-20 & 05 38 41.30 & -69 02 58.20 & 12.62 $\pm$ 0.02 & 11.86 $\pm$ 0.03 & 10.60 $\pm$ 0.02 &  8.52 $\pm$ 0.02 &    I \\

30Dor-21 & 05 38 43.20 & -69 06 59.80 & 10.54 $\pm$ 0.02 &  9.67 $\pm$ 0.03 &  8.75 $\pm$ 0.02 &  6.52 $\pm$ 0.02 &    I\tablenotemark{*}\\

30Dor-22 & 05 38 44.30 & -69 06 05.80 & 10.89 $\pm$ 0.02 & 10.72 $\pm$ 0.03 & 10.50 $\pm$ 0.02 &  8.44 $\pm$ 0.02 & I/II\tablenotemark{*}\\

30Dor-23 & 05 38 45.30 & -69 04 41.50 & 11.08 $\pm$ 0.02 & 10.91 $\pm$ 0.03 &  8.33 $\pm$ 0.02 &  6.62 $\pm$ 0.02 & I/II\tablenotemark{*}\\

30Dor-24 & 05 38 45.40 & -69 02 51.30 & 13.87 $\pm$ 0.02 & 13.43 $\pm$ 0.04 & 11.06 $\pm$ 0.02 &  9.00 $\pm$ 0.02 &    I \\

30Dor-25 & 05 38 48.80 & -69 01 38.60 &  9.90 $\pm$ 0.02 &  9.59 $\pm$ 0.03 &  9.45 $\pm$ 0.02 &  9.00 $\pm$ 0.02 &   II \\

30Dor-26 & 05 38 48.80 & -68 58 39.00 & 13.63 $\pm$ 0.02 & 13.69 $\pm$ 0.03 & 12.01 $\pm$ 0.02 & 10.60 $\pm$ 0.02 & I/II \\

30Dor-27 & 05 38 49.80 & -69 06 42.80 & 10.64 $\pm$ 0.02 &  9.81 $\pm$ 0.03 &  9.04 $\pm$ 0.02 &  6.84 $\pm$ 0.02 &    I\tablenotemark{*}\\

30Dor-28 & 05 38 51.20 & -69 06 41.00 &  9.87 $\pm$ 0.02 &  9.74 $\pm$ 0.03 &  9.18 $\pm$ 0.02 &  7.32 $\pm$ 0.02 & I/II \\

30Dor-29 & 05 38 53.90 & -69 09 31.10 & 11.57 $\pm$ 0.02 & 11.33 $\pm$ 0.03 &  8.94 $\pm$ 0.02 &  7.28 $\pm$ 0.02 & I/II \\

30Dor-30 & 05 38 56.40 & -69 04 16.10 & 10.95 $\pm$ 0.02 &  9.01 $\pm$ 0.03 &  7.52 $\pm$ 0.02 &  6.25 $\pm$ 0.02 &    I\tablenotemark{*}\\

30Dor-31 & 05 38 56.90 & -69 07 30.70 & 12.48 $\pm$ 0.02 & 11.87 $\pm$ 0.03 & 10.26 $\pm$ 0.02 &  8.41 $\pm$ 0.02 &    I \\

30Dor-32 & 05 38 57.30 & -69 07 09.50 & 10.68 $\pm$ 0.02 & 10.22 $\pm$ 0.03 & 10.02 $\pm$ 0.02 &  9.62 $\pm$ 0.02 &   II \\

30Dor-33 & 05 38 58.50 & -69 08 42.10 & 12.42 $\pm$ 0.02 & 12.42 $\pm$ 0.03 &  9.63 $\pm$ 0.02 &  7.88 $\pm$ 0.02 & I/II \\

30Dor-34 & 05 38 59.50 & -69 05 08.60 & 11.05 $\pm$ 0.02 & 10.18 $\pm$ 0.03 &  9.66 $\pm$ 0.02 &  7.69 $\pm$ 0.02 &    I \\

30Dor-35 & 05 39 00.50 & -69 08 41.00 & 12.28 $\pm$ 0.02 & 12.42 $\pm$ 0.03 & 10.13 $\pm$ 0.02 &  8.48 $\pm$ 0.02 & I/II \\

30Dor-36 & 05 39 04.50 & -69 04 13.80 & 11.71 $\pm$ 0.02 & 10.98 $\pm$ 0.03 &  9.20 $\pm$ 0.02 &  7.71 $\pm$ 0.02 &    I \\

30Dor-37 & 05 39 29.20 & -69 13 26.60 &  9.58 $\pm$ 0.02 &  9.17 $\pm$ 0.03 &  8.76 $\pm$ 0.02 &  8.34 $\pm$ 0.02 &   II \\

30Dor-38 & 05 39 35.20 & -69 04 00.40 & 12.59 $\pm$ 0.02 & 13.16 $\pm$ 0.03 & 10.60 $\pm$ 0.02 &  8.92 $\pm$ 0.02 & I/II \\

30Dor-39 & 05 39 38.30 & -68 57 39.90 & 13.93 $\pm$ 0.02 & 13.84 $\pm$ 0.03 & 12.68 $\pm$ 0.02 & 10.26 $\pm$ 0.02 & I/II \\

30Dor-40 & 05 39 38.50 & -69 09 00.30 & 10.56 $\pm$ 0.02 &  9.78 $\pm$ 0.03 &  9.26 $\pm$ 0.02 &  8.64 $\pm$ 0.02 &   II \\

30Dor-41 & 05 39 39.40 & -69 11 51.80 &  8.11 $\pm$ 0.02 &  7.88 $\pm$ 0.03 &  7.49 $\pm$ 0.02 &  6.98 $\pm$ 0.02 &   II \\
\enddata
\newline
\tablenotetext{*}{These sources correspond to the YSO candidates studied by Brandner et al. (2001).}
\end{deluxetable}
\newpage
%Figure 1: spectra
\begin{figure}
\figurenum{1}
\plotone{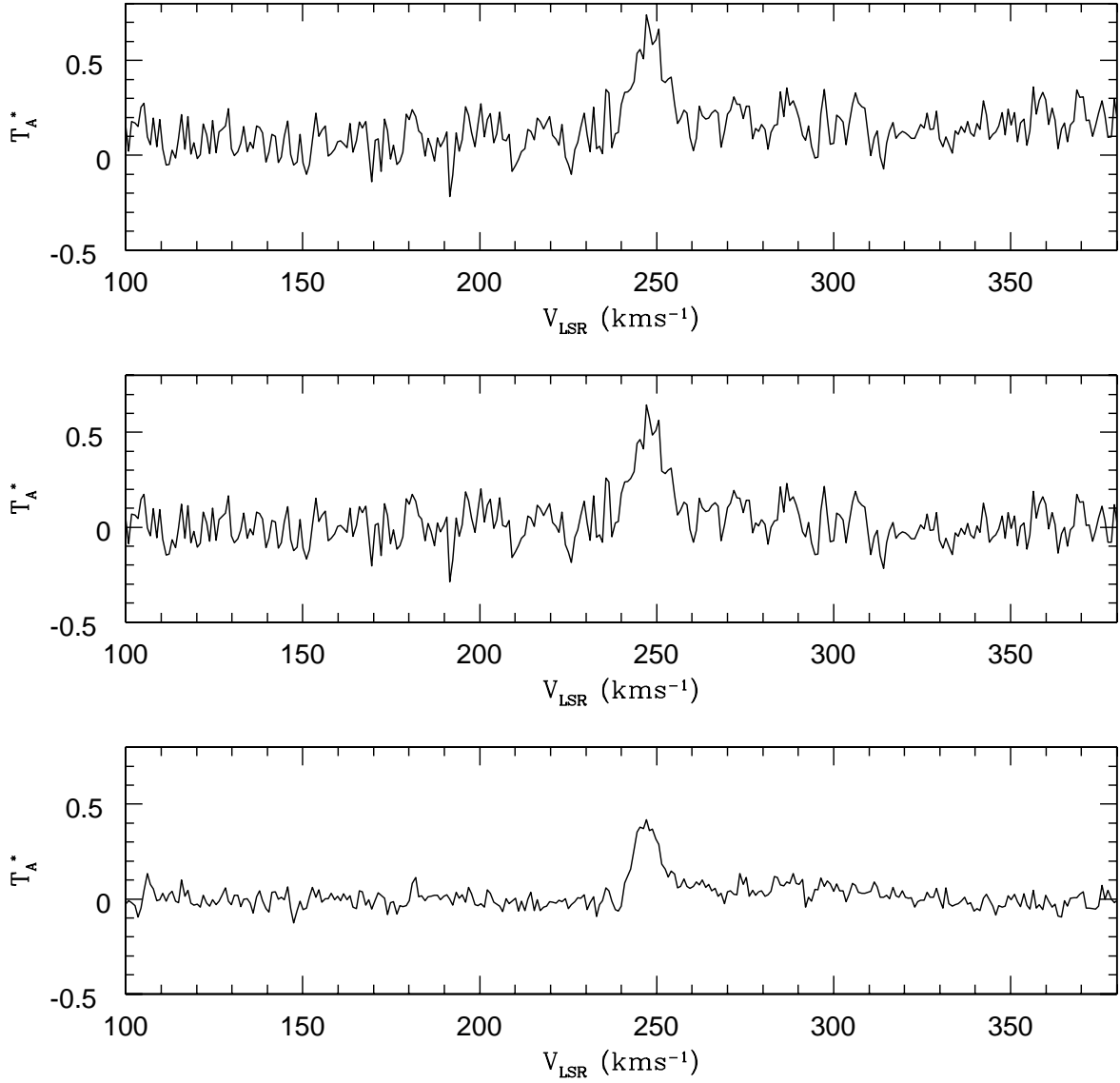}
\caption{The $^{12}$CO $J=2\rightarrow1$ raw spectra ($Top$: integration of 1 min and 30 sec), Fourier Transformed and baseline removed spectra ($Middle$), and integrated for 7 min and 20 sec spectra ($Bottom$) observed towards the peak of $^{12}$CO $J=4\rightarrow3$ emission in the 30 Doradus complex.}
\end{figure}

%figure 2 : Halpha image + co

\begin{figure}
\figurenum{2} \plottwo{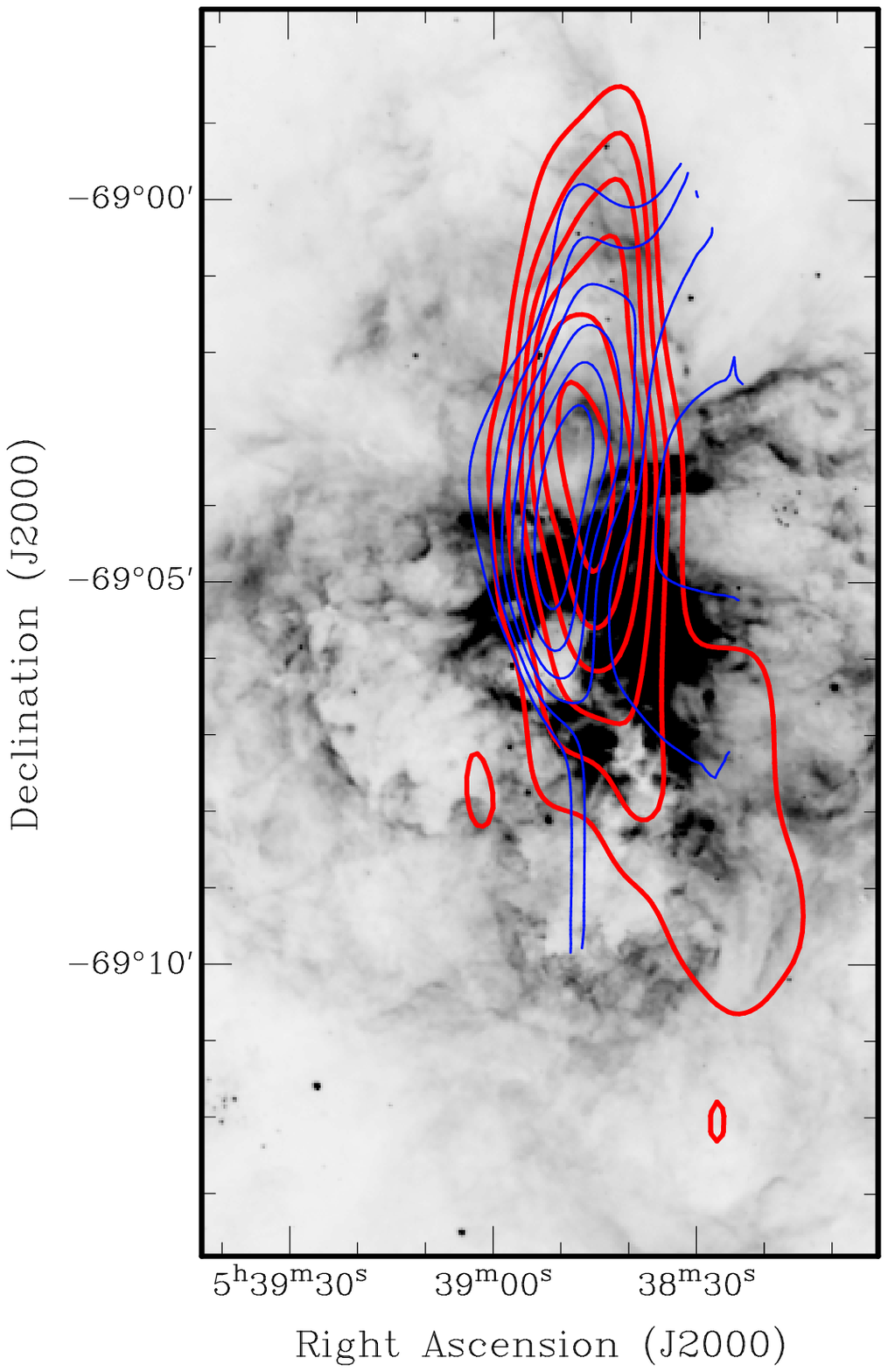}{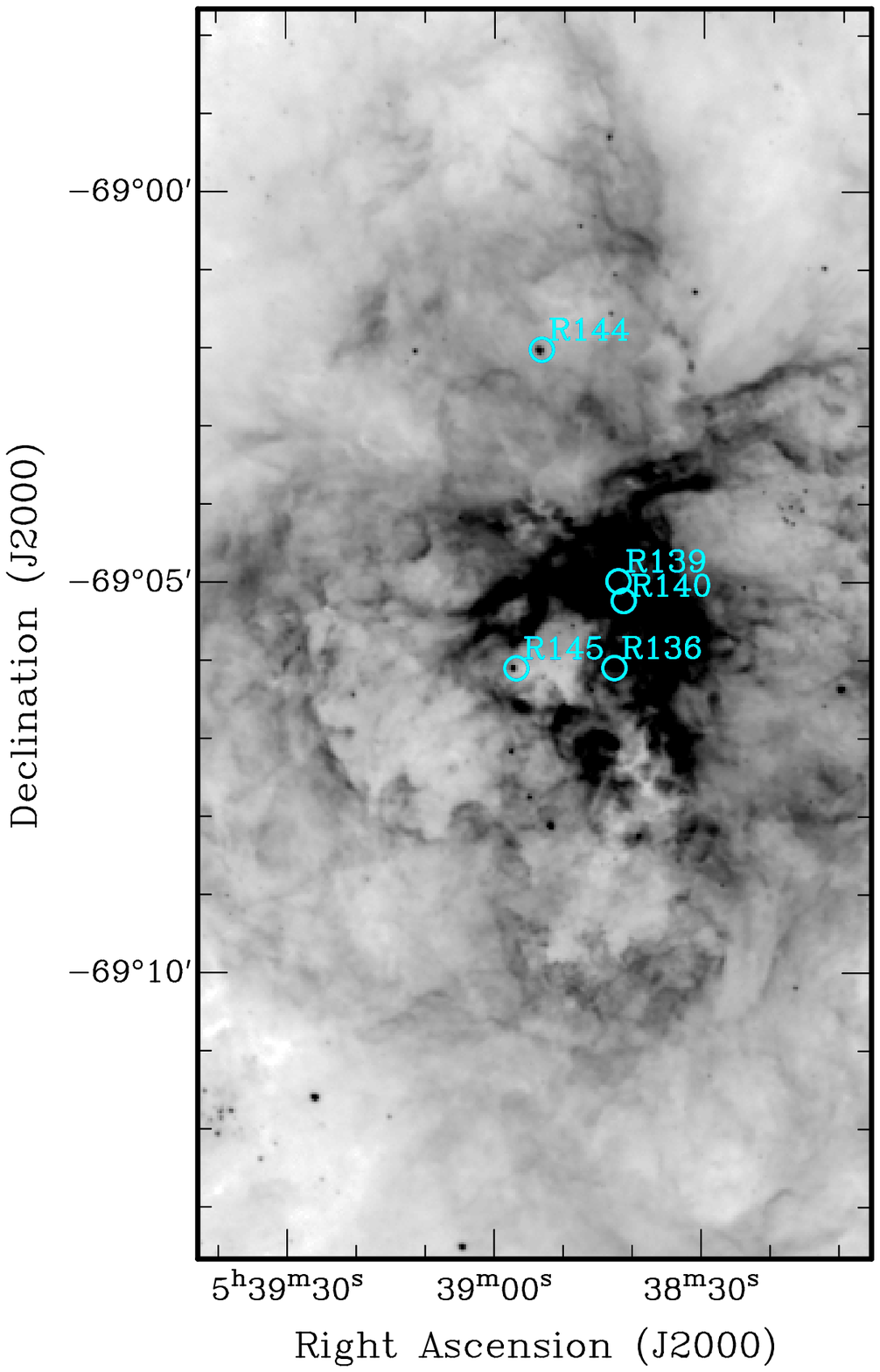}\caption{The $^{12}$CO $J=4\rightarrow3$ emission line (blue contour) is displayed and compared to the $^{12}$CO $J=2\rightarrow1$ emission line (red contour) from this position. These are overlayed on a grayscale image of H$\alpha$. The contour levels are 1.3, 1.7, 2.1, 2.5, 2.9, and 3.3 K km/s. H$\alpha$ image is taken with a CCD camera mounted at the Curtis Schmidt telescope at CTIO by C. Smith.}
\end{figure}

%figure 3 : xray + co
\begin{figure}
\figurenum{3} \plotone{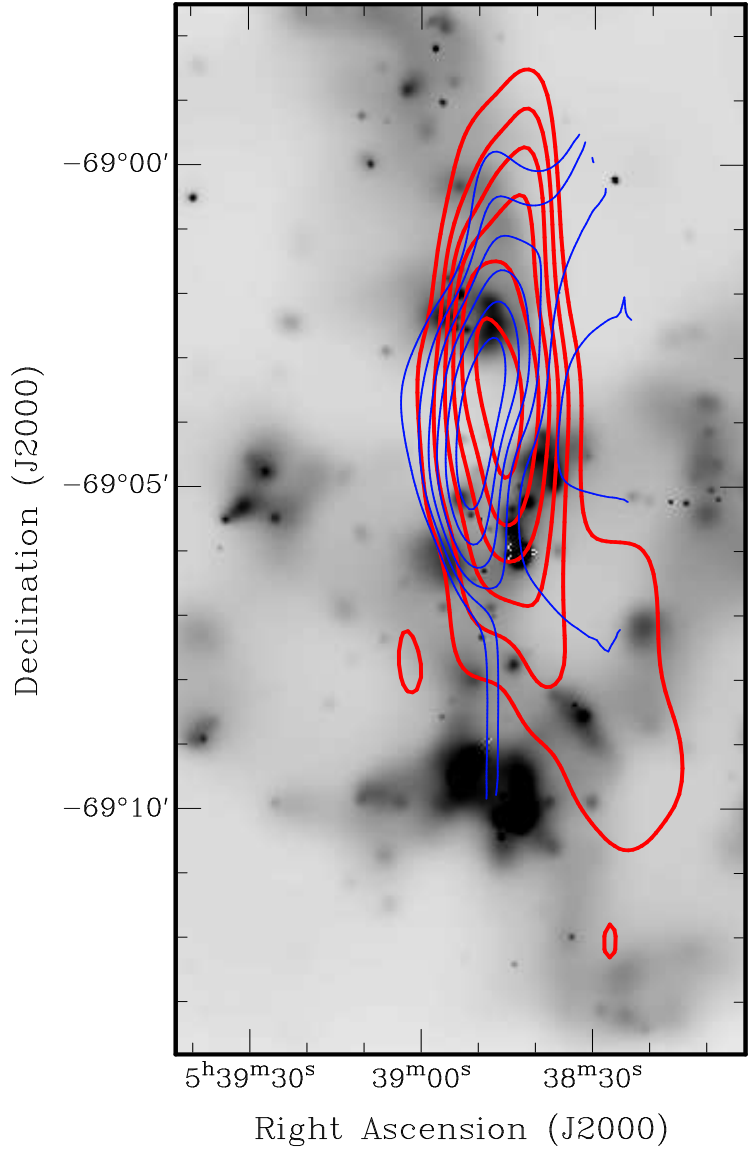} \caption{The $^{12}$CO $J=4\rightarrow3$
and $^{12}$CO$J=1\rightarrow0$ emission lines are overlayed on the Chandra X-ray image of 30 Doradus. Chandra ACIS-I 0.5-10 keV observations were exposure corrected and the resultant image was obtained from binning by 4 pixels and has angular resolution of 2''.}
\end{figure}

%figure 4 : spitzer image + co
\begin{figure}
\figurenum{4} \plotone{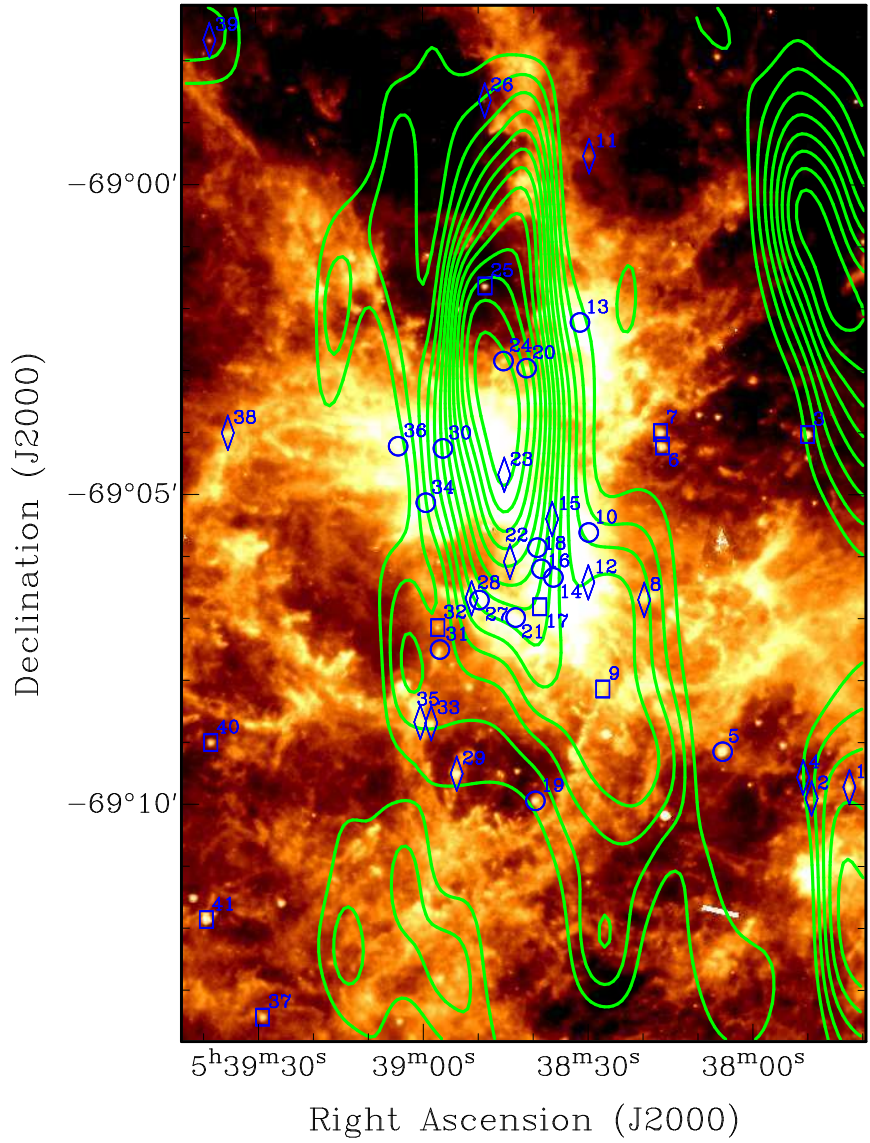} \caption{Molecular cloud core found
in the $^{12}$CO $J=2\rightarrow1$ emission map is overlaid on the
8.0$\mu$m $Spitzer$ image. The 41 YSO candidates are marked by the
same symbols used in Figure 5.}
\end{figure}

%figure 5 : color-color diagram

\begin{figure}
\figurenum{5} \plotone{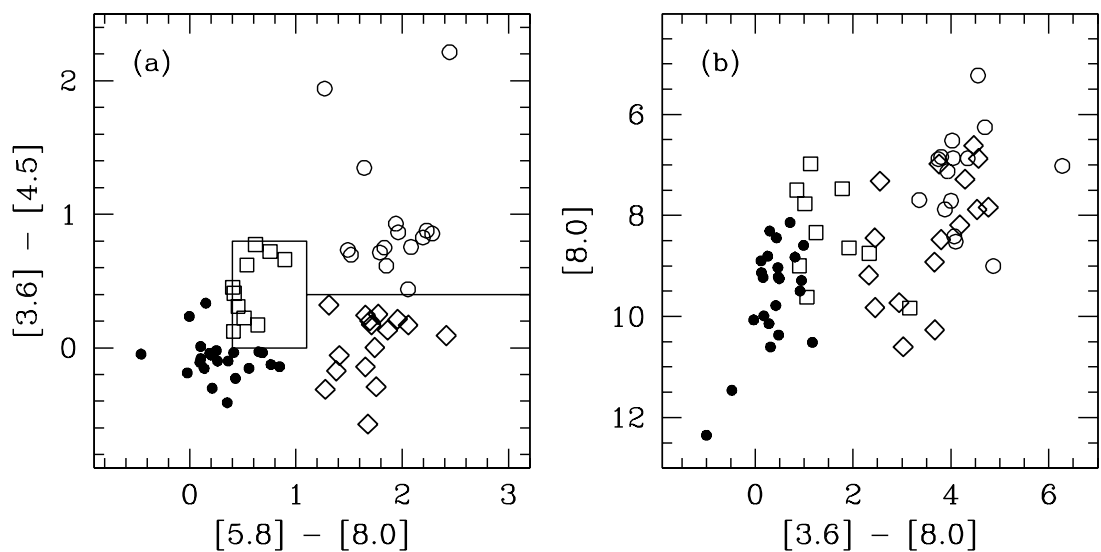} \caption{(a) Color-color diagram of
3.6 $\mu$m - 4.5 $\mu$m versus 5.8 $\mu$m - 8.0 $\mu$m. Detected
infrared sources are classified based on the criterion suggested by
Allen et al. (2004) and Megeath et al. (2004). Open circles
represent Class I objects and diamonds represent Class
\textrm{I}/\textrm{II} objects. Squares represent Class \textrm{II}
objects and Filled circles represent both Class \textrm{III} objects
and ordinary stars. Solid box represents the region of Class II sources
(Allen et al. 2004). The solid line shows the division between the Class
I and Class I/II candidates. (b) Color-magnitude diagram of 3.6 $\mu$m
- 8.0 $\mu$m versus 8.0 $\mu$m using the same symbols.}
\end{figure}
%figure 6 : color-ccolor, no 8um
\begin{figure}
\figurenum{6} \plotone{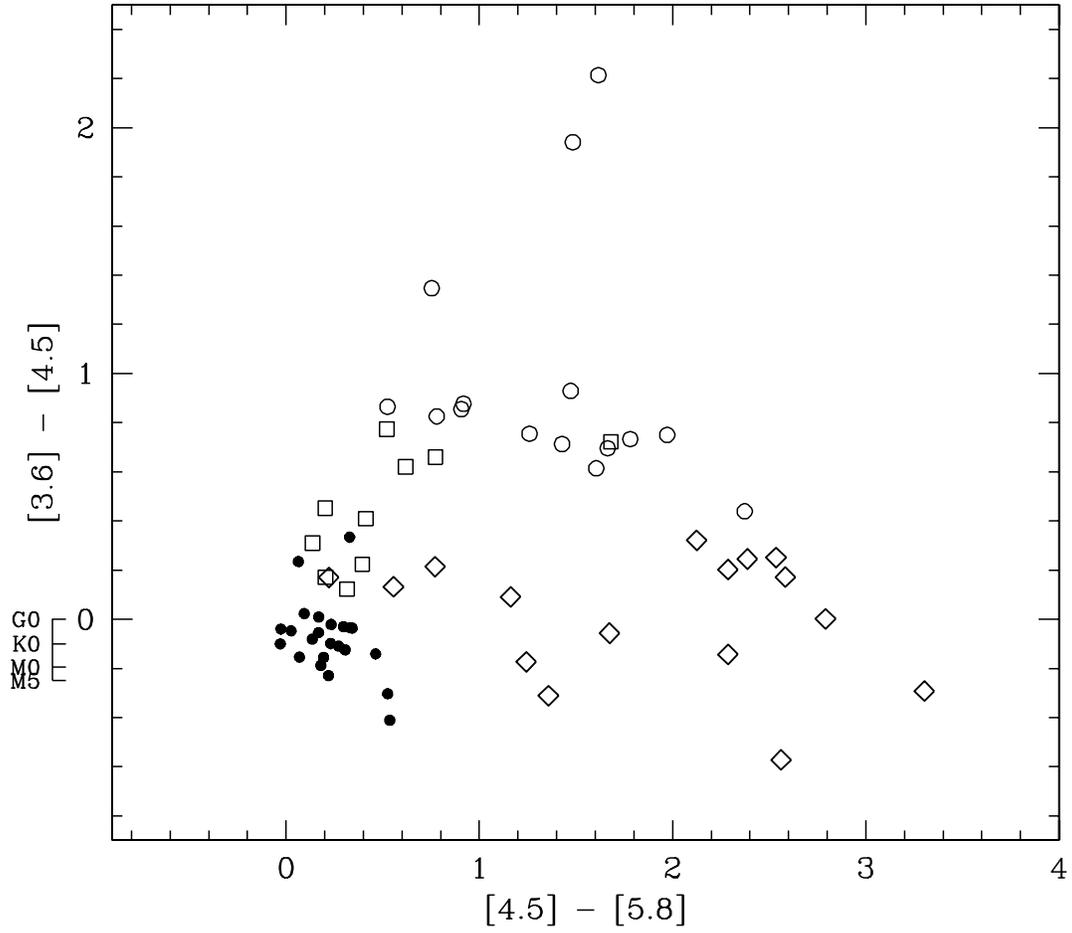} \caption{Color-Color Diagram of
3.6 $\mu$m - 4.5 $\mu$m versus 4.5 $\mu$m - 5.8 $\mu$m. The 41 YSO
candidates are marked by the same symbols used in Figure 5. We plot the 
locations of the late spectral types from Jones et al. (2005) along the 
vertical axis.}
\end{figure}
%figure 7 : Class I/II/III
\begin{figure}
\figurenum{7} \plotone{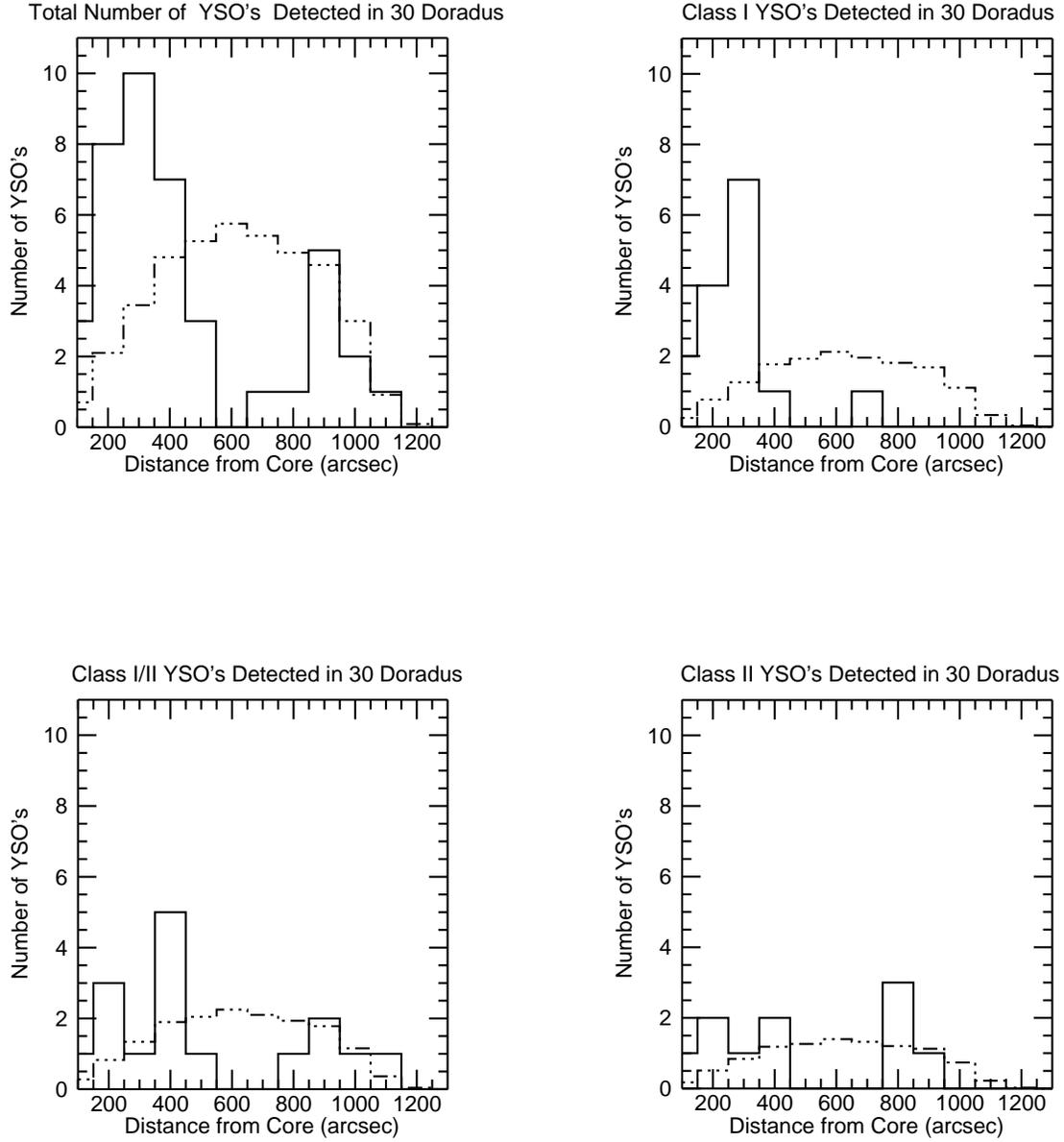}
\caption{The numbers of detected objects for each class are plotted
against the radial distance from the $^{12}$CO cloud core. The solid lines
are the objects observed with the IRAC/Spitzer data, while the dotted
lines represent the average values of the randomly generated fields.}
\end{figure}

\end{document}